\documentclass[11pt, reqno]{amsart}

\def\st{\tilde{s}}
\def\Pt{\tilde{{\cal P}}}
\def\Ct{\tilde{\CC}}
\def\gt{\tilde{g}}

\def\cal{\mathcal}

\def\a{\alpha}

\def\b{\beta}

\def\d{\delta}
\def\e{\varepsilon}
\def\ka{\kappa}
\def\l{\lambda}

\def\O{\Omega}

\def\CP1{{\mathbb C}P^1}
\def\Pcal{{\cal P}}

\def\B{{\bf B}}
\def\C{{\mathbb C}}
\def\Z{{\mathbb Z}}

\def\la{\label}

\def\c{\cite}
\def\f{\frac}

\def\p{\partial}

\def\rb{{\bf r}}

\def\0{S}
\def\log{\ln}
\def\ol{\overline}
\def\M{{\cal M}}
\def\Lh{\hat{\L}}

\def\B{{\bf B}}
\def\la{\label}

\def\c{\cite}
\def\f{\frac}

\def\p{\partial}

\def\0{S}
\def\1{T}

\def\log{\ln}
\def\ol{\overline}
\def\i{{\rm i}}

\def\Lh{{\widehat{{\cal L}}}}

\def\bar{\overline}
\def\det{{\rm det}}
\def\dbar{\bar{\partial}}

\def\Ct{\tilde{\CC}}
\def\ol{\overline}

\def\Mcal {{\mathcal M}}

\def \eqref #1{ (\ref{#1})}

\newlength{\dinwidth}
\newlength{\dinmargin}
\def \&{\hspace{-18pt}&}

\newtheorem{theorem}{Theorem}[section]

\newtheorem{corollary}[theorem]{Corollary}
\newtheorem{remark}[theorem]{Remark}

\def\be{\begin{equation}}
\def\ee{\end{equation}}
\def\ben{\begin{displaymath}}
\def\een{\end{displaymath}}
\def\baa{\begin{eqnarray}}
\def\eaa{\end{eqnarray}}

\def\ba{\begin{array}}
\def\ea{\end{array}}
\def\la{\label}
\def\p{\partial}
\def\nz{{\bf M}}

\def\Pcal{{\mathcal P}}
\def\Hcal{{\mathcal H}}
\def\rb{{\bf r}}

\def\Mb{{\mathfrak M}}
\def\pih{\hat{\pi}}

\def\ah{\hat{a}}
\def\bh{\hat{b}}

\def\Ch{\hat{C}}

\makeatletter
\@addtoreset{equation}{section}
\makeatother

\def\olMcal{\overline{\Mcal}}

\def\Lh{{\hat{\Lambda}}}
\def\lh{{\hat{\lambda}}}

\def\C{{\mathbb C}}
\def\R{{\mathbb R}}
\def\Z{{\mathbb Z}}
\def\Q{{\mathbb Q}}
\def\B{\Omega}
\def\O{\Omega}
\def\Oh{{\widehat{\Omega}}}

\def\Mcal{{\mathcal M}}
\def\Lcal{{\mathcal L}}
\def\M_g{{\mathcal{\M}_g}}

\def\Mcal{{\mathcal M}}
\def\f{\frac}
\def\e{\epsilon} 
\def\d{\delta}
\def\a{\alpha}
\def\b{\beta}

\def\deg{{\rm deg}}

\def\CC{\Sigma}
\def\i{\sqrt{-1}}

\def\Ch{\widehat{\Sigma}}
\def\Ccal{{\mathcal C}}

\def\gh{\hat{g}}

\def\ka{\kappa}

\def\Lcal{{\mathcal L}}
\def\Acal{{\mathcal A}}

\def\l{\lambda}

\def\Oh{\hat{\Omega}}
\def\gh{\hat{g}}

\def\tauh{\widehat{\tau}}

\def\dim{{\rm dim}}

\def\be{\begin{equation}}
\def\ee{\end{equation}}
\def\ben{\begin{displaymath}}
\def\een{\end{displaymath}}
\def\baa{\begin{eqnarray}}
\def\eaa{\end{eqnarray}}

\def\ba{\begin{array}}
\def\ea{\end{array}}
\def\la{\label}
\def\p{\partial}

\makeatletter
\@addtoreset{equation}{section}
\makeatother

\def\C{{\mathbb C}}
\def\R{{\mathbb R}}
\def\Z{{\mathbb Z}}
\def\Q{{\mathbb Q}}

\def\M{{\mathcal M}}

\def\2x2{{\left(\!\!\begin{array}{cc}a&b\\c&d\\\end{array}\!\!\right)}}

\def\f{\frac}
\def\e{\epsilon}

\def\d{\delta}
\def\a{\alpha}

\def\deg{{\rm deg}}

\def\B{\bf B}
\def\p{\partial}

\def\Lcal{{\mathcal L}}
\def\Ccal{{\mathcal C}}

\def\ah{\hat{a}}
\def\bh{\hat{b}}
\def\f{\frac}
\def\l{\lambda}

\def\Acal{{\mathcal A}}
\def\a{\alpha}
\def\b{\beta}
\def\p{\partial}
\def\e{\epsilon}

\def\la{\label}
\def\tauh{\hat{\tau}}
\def\dif{Q}

\usepackage{geometry}                
\usepackage{graphicx}
\usepackage{amssymb}
\usepackage{epstopdf}

\title[Tau functions and Hitchin's spectral covers]{Tau functions, Hodge classes and  discriminant loci on moduli spaces of Hitchin's spectral covers}
\author{Dmitry Korotkin}
\address{Department of Mathematics and Statistics, Concordia University, 1455 de Maisonneuve W., Montreal, Qu\'ebec, 
Canada H3G 1M8}
\email{dmitry.korotkin@concordia.ca}
\author{Peter Zograf}
\address{Steklov Mathematical Institute, Fontanka 27, Saint Petersburg 191023
Russia; Chebyshev Laboratory, Saint Petersburg State University, 14-th Line
V.O. 29, Saint Petersburg 199178 Russia}
\email{zograf@pdmi.ras.ru}

\begin{document}

\hskip7.5cm 
Dedicated to the memory of L.D.Faddeev
\vskip0.5cm
\maketitle

{\bf Abstract.}  
We define two tau functions,  $\tau$ and $\tauh$, on  moduli spaces of  spectral covers of $GL(n)$  Hitchin's systems.
Analyzing the properties of $\tau$, we  express the  divisor class of  the universal  Hitchin's discriminant
in terms of standard generators of the rational Picard group of the moduli spaces of spectral covers with variable base.
The function $\tauh$ is used to compute the divisor of canonical 1-forms with multiple zeros. 
\tableofcontents

\section{Introduction}

Yang-Mills equations have a deep connection to the theory of integrable systems: most of  soliton equations are dimensional reductions of the self-dual Yang-Mills equation (SDYM). In the pioneering work \cite{Hitchin1,Hitchin2}, N.~Hitchin  proposed a dimensional reduction of SDYM by
splitting 4-dimensional space into the product of a Riemann surface $\CC$ and the real plane $\R^2$, where the gauge fields are assumed to 
be independent of coordinates on $\R^2$. As a result of such dimensional reduction, one arrives at the class of finite-dimensional 
completely integrable systems, called {\it Hitchin's systems}; we refer to  Atiyah's book \cite{Atiyah} (Sect. 6.3) for an introduction to the topic and to the original papers  \cite{Hitchin1,Hitchin2} and  reviews \cite{Donagi,DonagiMarkman} for more detailed description of the subject.
Hitchin's  systems, as well as their generalizations to the meromorphic case \cite{Hurtub}, provide the most general class of integrable systems associated to Riemann surfaces of an arbitrary genus. 

Let $\CC$ be a Riemann surface (smooth complex curve) of genus $g\geq 1$. The Hamiltonians of Hitchin's system are encoded in the so-called {\it spectral cover}
$\Ch$ which is an $n$-sheeted cover of $\CC$ defined by the equation in $T^*\CC$
\be
\Ch=\{(x,v)\in\CC\times T^*_x\CC\;|\; P_n(v)=0\}
\la{cancov}
\ee
where 
$$P_n(v)=v^n+\dif_1v^{n-1}+\dots+\dif_{n-1}v+\dif_n,$$
$\dif_k$ is a holomorphic $k$-differential on $\CC$, and $v$ is considered as a holomorphic 1-form on 
 $\Ch$.  In the framework of \cite{Hitchin1} the equation (\ref{cancov}) is given by the characteristic polynomial
 $P_n(v )={\rm det}(\Phi-vI)$ 
  of the so-calles {\it Higgs field} $\Phi$ on $\CC$.

For the most general case of $GL(n)=GL(n,\C)$ Hitchin's systems all differentials $\dif_k$ from  (\ref{cancov}) are arbitrary;
in the case of $SL(n)$ systems  $\dif_1=0$. In this paper we mainly consider the
 generic  case of
$GL(n)$ systems although most of the formulas are applicable to the $SL(n)=SL(n,\C)$ case without modification.

The branch points of the cover $\Ch$ are the zeros of the discriminant $W$ of $P_n(v)$ that coincides with the resultant of $P_n$ and $P_n'$
up to a sign:
\be
W={\rm Discr}\,(P_n)=(-1)^{\f{n(n-1)}{2}}{\rm Res}\,(P_n,P_n')\;.
\la{discr}\ee
It is easy to verify that the discriminant $W$ is a holomorphic $n(n-1)$-differential on $\CC$. Thus, the number $m$ of zeros of $W$, counted with multiplicities, equals 
\be
m=n(n-1)(2g-2)\;,
\ee
and the Riemann-Hurwitz formula 
gives  the genus $\gh$ of $\Ch$:
\be
\gh=n^2(g-1)+1\;
\la{gh}
\ee
so that $m=2(\gh-1-n(g-1))$.
When all zeros of $W$ are simple, all branch points of the cover $\pih:\Ch\to \CC$ are also simple.
In the simplest case of (\ref{cancov})  $n=2$; in particular, for
 $SL(2)$ we have $\dif_1=0$ and the equation (\ref{cancov}) takes the form
\be
v^2+\dif_2=0\,,
\la{can2}
\ee 
where $\dif_2$ is a holomorphic quadratic differential. The cover defined by (\ref{can2}) is sometimes called a {\em ``canonical cover"} \cite{Abikoff,DouadyHubbard}. The genus of (\ref{can2}) equals $4g-3$ (assuming that all zeros of $\dif_2$ 
are simple) and the dimension of the moduli space of curves (\ref{can2}) equals $6g-6$ when the base curve $\CC$ is also allowed to vary. Since the space of covers (\ref{can2}) forms an open subspace in the cotangent bundle $T^*\M_g$ of the moduli space $\M_g$ of curves of genus $g$, it possesses a canonical symplectic structure. This
symplectic structure, including a natural system of {\it period}, or {\it homological}, Darboux coordinates was studied
in detail in the recent paper \cite{BKN}.

An immediate  generalization of (\ref{can2}) is given by the family of $\Z_n$-invariant covers
\be
v^n+\dif_n=0\;,
\la{cann}
\ee
where $\dif_n$ is a holomorphic $n$-differential and $\Z_n=\Z/n\Z$. The genus of the covering (\ref{cann}) is also given by (\ref{gh}), but this case is
far from being generic since all ramification points of (\ref{cann}) are of order $n$.
The moduli space of $\Z_n$-covers (\ref{cann}) was studied in \cite{KSZ}, see also \cite{GruMol}.

The goal of this paper is to extend some of the results  about the moduli spaces of $\Z_n$-covers  
to the moduli spaces of Hitchin's generic $GL(n)$ covers (\ref{cancov}). 

In particular, we generalize the theory of tau functions (which can be considered as a higher genus generalizations of Dedekind's eta function) to the moduli spaces of Hitchin's covers. 
For moduli of $\Z_n$-curves this was done in \cite{contemp} (for $n=2$)  and in \cite{KSZ} (for $n>2$),
using the approach developed earlier in \cite{JDG, Advances,MRL}.

 In particular, the Bergman tau functions (called so due to their close ties to the Bergman projective connection) allowed to find new relations in the Picard groups of these moduli spaces. 

We also notice that the tau functions we discuss here can be interpreted as determinants of appropriate $\dbar$-operators 
in the spirit of \cite{Quillen}; they also have close relations to conformal field theory \cite{Kniz87}, isomonodromic deformations
\cite{JimboMiwa,Malg83,KitKor,Annalen} and Frobenius manifolds \cite{Dubr1,Dubr2,IMRN1,IMRN2}.



{\bf Spaces of coverings with fixed base.}
Let $\CC$ be a smooth curve of genus $g$ and denote by  $\Mcal^{\CC}$ the moduli space of $GL(n)$ spectral covers of the form  (\ref{cancov}). Then
\be
\Mcal^{\CC}=\bigoplus_{j=1}^n H^0(\CC,K_{\CC}^{\otimes j})\,\label{wpr}
\ee
where $K_{\CC}=T^*\CC$ is the canonical line bundle on $\CC$,
and
\be
\dim\,\M^{\CC}=\gh=n^2(g-1)+1
\la{dimMC}
\ee
(recall that $\dim H^{0}(\CC,K)=g$ and $ \dim H^{0}(\CC,K^{\otimes j})=(2j-1)(g-1)$  for $j\geq 2$).

There is a natural coordinate system on $\Mcal^{\CC}$ given by the $a$-periods of $v$:
\be
\Pcal_j=\int_{\ah_j}v
\la{coordM0}
\ee
where  
$\{\ah_j,\bh_j\}_{j=1}^{\gh}$ is a canonical symplectic basis in $H_1(\Ch,\Z)$.

We consider the following two codimension 1 loci in $\M^{\CC}$ -- the locus $D_W^{\CC}$ of sets $\dif_1,\ldots,\dif_n$ of differentials such that the discriminant
$W$ of $P_n(v)$ has multiple zeroes, and the locus $D_v^\Sigma$ of sets $\dif_1,\ldots,\dif_n$ such that the Abelian differential $v$ on $\Ch$ has multiple zeroes. The locus $D_W^{\CC}$ is called the {\it Hitchin's discriminant locus}, whereas we call $D_v^\Sigma$ the locus of {\em degenerate spectral covers}.
For a generic point in $\M^{\CC}$, that is a point in the complement $\M^{\CC}\setminus (D_W^{\CC}\cup D_v^{\CC})$, all zeros of the discriminant $W$ and all zeros of the Abelian differential $v$ on $\Ch$ are simple. 

Consider also the space $\Mcal_W^{\CC}=\Mcal^{\CC}\setminus D_W^{\CC}$ of spectral covers with simple branch points and the space 
$\Mcal_v^{\CC}=\Mcal^{\CC}\setminus D_v^{\CC}$ of covers with simple zeros of $v$.


{\bf Spaces of covers with variable base.} 
Let $\ol{\M}_g$ be the Deligne-Mumford compactification of the moduli space of curves $\M_g$, let 
$\nu: \overline{\mathcal{C}}_g \to \ol{\M}_g$ be the universal curve, and let 
$\omega_g=\omega_{\overline{\mathcal{C}}_g/\ol{\M}_g}$ be the relative dualizing sheaf. Put
\be
\ol{\Mcal}=\bigoplus_{j=1}^n\Omega_g^{(j)}\;,
\ee
where $\Omega_g^{(j)}=R^0\nu_*\omega_g^{\otimes j}$ is the direct image of the $j$th power of $\omega_g$.
We have 
\be
\dim\,\Mcal=\dim\,\Mcal^{\CC}+3g-3=(n^2+3)(g-1)+1
\la{dimMcal}
\ee
There is also a natural forgetful map $\ol{\Mcal}\to \ol{\M}_g$ such that the fiber over $\CC\in M_g$ coincides with $\Mcal^\CC$ (the fibers over nodal curves are described in detail in \cite{KSZ}). Denote by $D_W$ and $D_v$ respectively  the unions of loci $D_W^{\CC}$ and $D_v^{\CC}$ as $\CC$ varies over the entire moduli space $\ol{\M}_g$.

There is a natural action of $\C^*$ on $\ol{\Mcal}$ that fiberwise looks like $\dif_j\mapsto\epsilon^j\dif_j,\;\epsilon\in\C^*,\;j=1,\ldots,n,$ and respects the codimension one loci $D_W$ and $D_v$. After projectivization both $P{D_W}$ and $PD_v$ become divisors in $P\ol{\Mcal}$.
There is a natural forgetful map
\be
h\;:\;P\ol{\Mcal}\longrightarrow\ol{\M}_g\;.
\ee
The main goal of this paper is to express the class of divisor $[P D_W]$ 
in terms of the standard generators of the rational Picard group 
${\rm Pic}(P\ol{\Mcal})\otimes\Q$.
We also express the class $[P D_v]$ via the natural divisorial classes on $P\ol{\Mcal}$.

\subsection{Components of the universal discriminant locus}
Assume that two simple zeros  $x_1$ and $x_2$ of $W$ coalesce to a double zero on $\CC$. To describe possible 
deformations of the cover $\Ch$, choose a system of generators $\{\alpha_i,\beta_i\}_{i=1}^g,\; \{\gamma_i\}_{i=1}^m$ of $\pi_1(\CC\setminus\{x_j\}_{j=1}^m, x_0)$
satisfying the standard relation
\be
 \prod_{i=1}^g \a_i\beta_i\a_i^{-1}\beta_i^{-1}\,\gamma_1\dots \gamma_m= id\;.
\ee
The covering $\Ch$ defines the group homomorphism $G:\pi_1(\CC\setminus\{x_j\}_{j=1}^m) \to S_n$, the permutation group of $n$ elements. Let $s_1=G(\gamma_1)$ and  $s_2=G(\gamma_2)$. 
When all zeros of $W$ (i.e. branch points of the covering $\Ch$) are simple, both $s_1$ and $s_2$ are simple permutations.
As $x_2\to x_1$, the covering $\Ch$ degenerates to a covering $\Ch_0$ whose structure depends on the type of the
product $s_1s_2$. 

Consider a neighborhood $U$ of $\CC$ containing both $x_1$ and $x_2$, and 
introduce a local coordinate $z$in $U$.
Let $z_1=z(x_1)$ and $z_2=z(x_2)$. There are three patterns of local behavior of $x_1$ and $x_2$  that correspond to three different components of $D_W$.
We will use the terminology of \cite{LZ}:
  
{\it 1. The ``boundary" $D_W^{(b)}$}. 

In this case $s_1s_2$ is a trivial permutation. In the limit $z_1\to z_2$
the spectral cover $\Ch$ acquires a node (double point) while approaching
the (Deligne-Mumford) boundary of $M_{\gh}$.  
Since the 
order of points $x_1$ and $x_2$ is irrelevant, a transversal local coordinate on $\Mcal$ near $D_W^{(b)}$
can be chosen as
\be
t_b = (z_1-z_2)^2\;.
\la{locb}
\ee
{\it 2. The ``Maxwell stratum" $D_W^{(m)}$}.

In this case $s_1s_2$ is a product of two cycles of length 2, i.e. the ramification points of $\Ch$ remain simple, but two of them correspond to the same critical value $x_1\in \CC$.
Then, since a point in the Maxwell stratum 
splits into two simple critical values in two ways, a transversal local coordinate on $\Mcal$ near $D_W^{(m)}$ can be chosen as
\be
t_m = z_1-z_2\;.
\la{locm}
\ee

 {\it 3. The ``caustic" $D_W^{(c)}$}.
 
In this case $s_1s_2$ is a cycle of length 3, i.e. as $x_2\to x_1$ the cover $\Ch$ acquires a ramification point of order 3. It can be decomposed into a product of two transpositions in 3 different ways, 
a transversal local coordinate on $\Mcal$ near 
 $D_W^{(c)}$ can be chosen as
\be
t_c = (z_1-z_2)^{2/3}\;.
\la{locc}
\ee

The transversal  local coordinates $t_b$, $t_m$ and $t_c$  can be specified further as follows.
Let us choose the coordinate $z\in U$ in such a way that the discriminant $W$ is given by
\be
W=(z-z_1)(z-z_2)\,dz^{n(n-1)}\;.
\la{WOm}
\ee
Put $w=W^{\frac{1}{n(n-1)}}$; then, up to a multiplicative constant,
$$
\int_{x_1}^{x_2} w \sim (z_1-z_2)^{\f{n(n-1)+2}{n(n-1)}}
$$
and
\be
z_1-z_2=const\cdot \left(\int_{x_1}^{x_2} w\right)^{\f{n(n-1)}{n(n-1)+2}}\;.
\la{zzw}
\ee


As one can see from the above considerations, the universal Hitchin's discriminant locus $D_W$ splits into 3 components:
\be
[D_W]= [D_W^{(b)}]+2[D_W^{(m)}]+3[D_W^{(c)}]\;.
\la{DWdec}
\ee
This splitting respects the action of $\C^*$ on $\ol{\Mcal}$ and descends to the projectivizations of these divisors.

\section{Tau functions on spaces of Abelian and higher order differentials}

Here we summarize previously known results from \cite{JDG,MRL,contemp,KSZ}.

\subsection{Preliminaries}

For a Torelli marked Riemann surface $\CC$ of genus $g$ introduce the canonical bidifferential $B(x,y)$, $x,y\in\CC,$
which has the quadratic pole with biresidue 1 on the diagonal and vanishing $a$-periods. 
The bidifferential $B$ is expressed via the the prime-form $E(x,y)$ as follows: $B(x,y)=d_x d_y {\rm log} E(x,y)$ (see \cite{Fay73,Tata} for details). Consider a basis of holomorphic differentials
$v_i$ on $\CC$ normalized by
$
\int_{a_\a}v_{\beta}=\delta_{\a\b}\;.
$
The period matrix $\O$ of $\CC$ is given by:
$
\O_{ij}=\int_{b_i} v_j\;.
$

In a local coordinate $\xi$ near the diagonal $\{x=y\}\subset \CC\times \CC$,  the bidifferential $B(x,y)$ has the expansion
\be
B(x,y)=\left(\f{1}{(\xi(x)-\xi(y))^2}+\f{S_B(\xi(x))}{6}
+O((\xi(x)-\xi(y))^2)\right)d\xi(x) d\xi(y),
\la{asW}
\ee
where $S_B$ is a projective connection on $\CC$ called the {\it Bergman projective connection}.

If two canonical bases of cycles on $\CC$, $\{a_\a',b_\a'\}_{\a=1}^g$ and 
$\{a_\a,b_\a\}_{\a=1}^g$
 are related by
a matrix 
\be
\sigma=
\left(\begin{array}{cc} D & C\\B & A \end{array}\right)\in Sp(2g,\Z)\;,
\la{symtrans}
\ee
then the corresponding canonical bidifferentials are related as follows (p. 21 of \cite{Fay73}):
\be
B^{\sigma}(x,y)=B(x,y)-2\pi \sqrt{-1}\sum_{i,j=1}^g  (C\O+D)^{-1}_{ij}\; v_i(x) v_j(y)\;.
\la{BsB}
\ee

The Abel map is defined by 
$ \Acal_{x_0}^i(x)=\int_{x_0}^x v_i $. 
Let us also define
\be
\Ccal(x)=\frac{1}{{\mathcal W}(x)}\left(\sum_{i=1}^g\,v_i(x)\frac{\partial}{\partial v}\right)^g\,\Theta(v;\O)
\left |_{v=K^x}\right.\;,
\la{Ccal}\ee
where
\be
{\mathcal W}(x)= {\rm \det} \left(v_i^{(j-1)}(x)\right)_{i,j=1}^g
\la{Wronks}
\ee
is the Wronskian determinant of the basic holomorphic differentials, $\Theta$ is the theta function and $K^x$ is the vector of Riemann constants with base point $x$.
The expression (\ref{Ccal}) is a multi-valued $g(1-g)/2$-differential on $\CC$ which  does not have any zeros or poles \cite{Fay92}.
In the case of  genus $1$ the $x$-dependence in (\ref{Ccal}) drops out and $\Ccal(x)$ turns into $\Theta'((\O+1)/2)$.

\subsection{Spaces of holomorphic Abelian differentials}

Denote by $\Hcal_g$ the moduli space of pairs $(\CC,v)$ where $\CC$ is a Riemann surface of genus $g$ and $v$ is a  holomorphic differential on $\CC$; clearly $\dim \Hcal_g=4g-3$.
The space $\Hcal_g$ can be stratified according to multiplicities of zeros of $v$: for any partition $[k_1,\dots,k_M]$ of $2g-2$ 
 denote by $\Hcal_g(k_1,\dots,k_M)$ the moduli space of pairs such that multiplicities of zeros $y_1,\dots,y_M$ of $v$ are  given by $\{k_i\}_{i=1}^M$.
Then  $\dim \Hcal_g(k_1,\dots,k_M)=2g+M-1$ and a system of period, or {\it homological},  coordinates on $\Hcal_g(k_1,\dots,k_M)$
can be obtained by integrating $v$ over a system of generators in the relative homology group $H_1(\CC,\{y_i\}_{i=1}^{\M})$,
see \cite{KonZor} for details. A natural choice of generators in this homology group is
\be
\{s_1,\dots,s_{2g+M-1}\}=\{a_1,\dots,a_g,b_1,\dots,b_g,l_2,\dots,l_M)
\la{sj}
\ee
where $\{a_j,b_j\}_{j=1}^g$ is a canonical basis of cycles on $\CC$ and $l_j$ is a path connecting the $y_1$ with $y_j$.

Periods of $v$ along the cycles (\ref{sj}) give a system of local coordinates on the stratum $\Hcal_g(k_1,\dots,k_M)$:
\be
\Pcal_{s_i}=\int_{s_i} v\;,\hskip0.7cm i=1,\dots,2g+M-1\;.
\la{perco}
\ee

The dual basis of cycles in $H_1(\CC\setminus\{y_i\}_{i=1}^{\M})$ is defined by
\be
\{s_1^*,\dots,s^*_{2g+M-1}\}=\{-b_1,\dots,-b_g,a_1,\dots,a_g, c_2,\dots,c_M)
\la{sjs}
\ee
 where $c_j$ is a small positively oriented circle around $y_j$, so that $s_i^*\circ s_j=\delta_{ij}$
(the symbol $\circ$ denotes here the intersection pairing of 1-cycles).

The differential $v$ gives rise to a natural coordinate on $\CC$. Pick a fundamental polygon $\CC_0$ of $\CC$
and put
\be
z(x)=\int_{y_1}^x v\;.
\la{flatco}
\ee
The (multivalued) coordinate $z$ is defined on $\CC$ everywhere except the zeros $y_i$; near $y_i$ the local coordinate, called {\it distinguished}, is given by 
\be
\zeta_i(x)=\left(\int_{y_i}^x v\right)^{1/(k_i+1)}\;.
\la{dist}
\ee

Tau functions on strata of moduli spaces of holomorphic abelian differentials were introduced in \cite{JDG}, by generalizing the notion
of isomonodromic Jimbo-Miwa tau function for Riemann-Hilbert problems \cite{Annalen,IMRN2}.
The tau function $\tau(\CC,v)$ is defined on the stratum   $\Hcal_g(k_1,\dots,k_M)$ by the system
\be
\f{\p\log\tau(\CC,v)}{\p \Pcal_{s_i}}=-\f{1}{2\pi \sqrt{-1}}\int_{s_i^*}\f{B^{reg}_v}{v} \;,\hskip0.7cm i=1,\dots,2g+M-1
\la{deftau}
\ee
where
\be
B^{reg}_v(x)=\left.\left(B(x,y)-\f{v(x)v(y)}{(\int_y^x v)^2}\right)\right|_{y=x}\;.
\la{Breg}
\ee

Introduce two vectors 
${\bf r}$ and ${\bf s}$ such that 
\be
\Acal_x((v))+2K^x+\Omega {\bf r}+{\bf s}=0\;.
\la{defZh}
\ee
Put
\be
E(x,y_i)=\lim_{y\to y_i} E(x,y) \sqrt{d\zeta_i(y)}\;,
\la{defEp}
\ee
and
\be
 E(y_i,y_j)=\lim_{x\to y_i, y\to y_j} E(x,y) \sqrt{d\zeta_i(x)} \sqrt{d\zeta_j(y)}
\la{defEpp}
\ee
where $\zeta_i$ is the distinguished local parameter (\ref{dist}) on $\CC$ near $y_i$.
Then the  solution of the system (\ref{deftau}) looks as follows (see \cite{JDG} for the proof):

\be
\tau(\CC,v)=\Ccal^{2/3}(x)\left(\f{v(x)}{\prod_{i=1}^{\nz}E^{k_i}(x,y_i)}\right)^{(g-1)/3} 
\left(\prod_{i<j} E(y_i,y_j)^{k_i k_j}\right)^{1/6}\!\!e^{-\f{\pi \i}{6} \langle\O {\rb},{\rb}\rangle-\f{2\pi \i}{3}\langle{\rb},K^x\rangle }
\la{taupfor}
\ee

Under the change of Torelli marking of $\CC$ given by symplectic matrix (\ref{symtrans})
$\tau(\CC,v)$ transforms as follows:
\be
\tau(\CC,v)\to \rho\, \det(C\O+D)\, \tau(\CC,v)\;,
\la{sym}
\ee
where $\rho$ is a root of unity of degree depending on the multiplicities $k_j$.

Another important property of the tau function is its behavior under the action of $\C^*$:
\be
\tau(\CC,\e v)=\e^{\f{1}{24}\sum_{j=1}^M \frac{  k_j(k_j+2)}{k_j+1}  } \tau(\CC,v)\;.
\la{hom}
\ee

The tau function can be used for obtaining relations between divisors in the rational Picard group of the strata
on the moduli space of Abelian differentials. 

For the main stratum $\Hcal_g(1,\dots,1)$ these relations 
were found in \cite{MRL}. Namely, let $P\Hcal_g(1,\dots,1)$ be the projectivization of $\Hcal_g(1,\dots,1)$ respect to the action of $\C^*$. Let $L$ be the 
tautological line bundle associated to the projection $\Hcal_g(1,\dots,1)\to P\Hcal_g(1,\dots,1)$, and denote by $\phi=c_1(L)$ its first Chern class. Furthermore, denote by  $\lambda$ the pullback to $P\Hcal_g(1,\dots,1)$ of the Hodge class on $\M_g$. 
Then we have the following relation in the rational
Picard group ${\rm Pic}(\overline{P\Hcal}_g(1,\dots,1))\otimes \Q$ of the compactification of $P\Hcal_g(1,\dots,1)$:
\be
\lambda=\f{g-1}{4}\phi+\f{1}{24}\d_\deg+\f{1}{12}\d_0+\f{1}{8}\sum_{j=1}^{[g/2]}\d_j\;.\label{mf}
\ee
Here $\d_\deg$ is the divisor of Abelian differentials with multiple zeroes, and $\d_j,\;j=0,\dots,[g/2],$ are the pullbacks of the classes of the Deligne-Mumford boundary divisors on $\M_g$; see \cite{MRL} for details. \footnote{This relation has later received pure algebraic proofs 
by D.~Zvonkine (unpublished) and by D.Chen \cite{Chen}}

\subsection{Spaces of holomorphic $N$-differentials}
\la{secZN}
 
The above result was extended further to the spaces of $N$-differentials in \cite{contemp} (for $N=2$) and \cite{KSZ} (for $N>2$).

Let $\Mb_g^{N}$ be the moduli space of equivalence classes of  pairs $(\CC,W)$ where $W$ is a holomorphic $N$-differential on $\CC$ (both  $\CC$ and $W$ 
are allowed to vary here).  We refer to \cite{KSZ,GruMol} for a precise definition of the space  $\Mb_g^{N}$ and its compactification $\overline{\Mb}_g^{N}$.

The dimension of $\Mb_g^{N}$ is the sum of $3g-3$ and $(2N-1)(g-1)$, i.e. 
\be
\dim \Mb_g^{N}=2(N+1)(g-1)\;.
\la{dimMb}
\ee

The space $\Mb_g^{N} $ has an open  subset $\Mb_g^{N,0}$, that consists of equivalence classes of pairs $(\CC,W)$,
where $\CC$ is a smooth curve, and $W$ has only simple zeroes.The complement $\overline{\Mb}_g^{N}\setminus \Mb_g^{N,0}$ is the union of $[g/2]+2$ divisors that we denote by 
$D_{deg}, D_0,\ldots, D_{[g/2]}$, where $D_{deg}$ is the divisor of {\it degenerate} $N$-differentials 
(i.e. having multiple zeroes), and $D_i\; (i=0,\ldots, [g/2])$ are the pullbacks of the components of the
Deligne-Mumford boundary of $\overline{\M}_g$.

A natural $\C^*$-action on $\overline{\Mb}_g^{N}$ is given by multiplication $W\to\epsilon W,\;\epsilon\in\C^*$.
Denote by $\Lcal$ the tautological line bundle associated with the canonical projection $\overline{\Mb}_g^{N}\to P \overline{\Mb}_g^{N}$ and put
$\psi=c_1(\Lcal)\in {\rm Pic}(P \overline{\Mb}_g^{N})\otimes\Q$.

Denote by $\lambda$ the Hodge class on $\ol{\Mb}^N_g$ (i.e. the pullback of the Hodge class from the moduli space of curves $\overline{\M}_g$),
and consider the classes of boundary divisors
$\delta_{i},\;i=0,\ldots,[g/2],$ in ${\rm Pic}(P \overline{\Mb}_g^{N})\otimes\Q$. Then the rational Picard group ${\rm Pic}(P \overline{\Mb}_g^{N})\otimes\Q$  is freely generated by the classes
$\psi, \lambda, \delta_0,\dots,\delta_{[g/2]}$ \cite{KSZ}.

To each pair $(\CC,W)$ one can naturally associate a canonical cyclic branched cover $p:\Ct\to \CC$ of degree $N$, where
\be
\Ct=\{(x,w)\in \CC\times T_x^*\CC |\;w^N=W\}\;.
\la{Nco}\ee

When all zeros of $W$ are simple, the cover $\Ct$ is smooth and its genus is $\gt=N^2(g-1)+1$. 
The cover $\Ct$ is invariant with respect to the natural $\Z/N\Z$-action $(x,w)\mapsto (x,\rho^k w)$ where 
$\rho=e^{2\pi\sqrt{-1}/N}$. Denote by $f:\Ct\to\Ct$ the automorphism of $\Ct$ corresponding to $k=1$.
By definition, the holomorphic 1-form $w$ satisfies  $f^*w=\rho w$.

The group $H_1(\Ct,\C)$ 
can  be decomposed into the eigenspaces of the automorphism $f_*$ 
\be
H_1(\Ct,\C)=\bigoplus_{k=0}^{N-1} {\cal S}_k\;,
\la{decH}\ee
where $\dim\, {\cal S}_0=2g$ and the dimensions of ${\cal S}_k$ are independent of $k$ and given by 
\be
\dim\, {\cal S}_k= (N+1)(2g-2)\;, \qquad k=1,\dots,N-1\;.
\la{dimHk}\ee

The differential $w$ has non-vanishing periods only over the cycles in ${\cal S}_1$; these periods can be used as local coordinates on 
the moduli space $\Mb_g^{N,0}$ \cite{KSZ,GruMol}:
\be
\Pt_i=\int_{\st_i}w\;,\qquad i=1,\dots,(N+1)(2g-2)\;.
\la{Ptco}
\ee
where 
\be
\st_1,\dots,\st_{(N+1)(2g-2)}
\la{stdef}\ee
 is a basis of of the eigenspace ${\cal S}_1$.

For any two cycles $s_1\in {\cal S}_{l}$ and $s_2\in {\cal S}_{k}$ we have $s_1\circ s_2=0$ unless $k+l=N$.
The spaces ${\cal S}_k$ and ${\cal S}_{N-k}$ are  therefore dual to each other with respect to the standard intersection pairing
(the space ${\cal S}_0$ can be identified with $H_1(\CC)$, and, therefore, it is dual to itself).

Therefore, one can introduce a set of cycles dual to (\ref{stdef}) which form a basis in the  space ${\cal S}_{N-1}$:
\be
\st_1^*,\dots,\st^*_{(N+1)(2g-2)},\qquad \st_i^*\circ \st_j=\delta_{ij}\;.
\la{stdefdu}\ee

Now assume that all zeros of $W$ are simple, i.e. 
\be
(W)=\sum_{i=1}^{N(2g-2)} x_i\;.
\la{divW2}\ee
Then 
the distinguished local coordinate on $\CC$ in a neighbourhood of the point $x_i$ is given by 
\be
\zeta_i(x)= \left(\int_{x_i}^x v\right)^{N/(N+1)}\;.
\la{distlp}
\ee
In terms of these coordinates we define
\begin{eqnarray*}
E(x,x_k)&=&\lim_{y\rightarrow x_k}E(x,\zeta(y))\sqrt{d\zeta_k(y)},\\
E(x_k,x_l)&=&\lim_{\stackrel{\scriptstyle x\rightarrow x_k}{y\rightarrow x_l}}E(x,y)
\sqrt{d\zeta_k(x)}\sqrt{d\zeta_l(y)}\,.
\end{eqnarray*}
We choose two vectors ${\bf r},\,{\bf s}\in\f{1}{n}\Z^{g}$ that satisfy the condition
\be
\f{1}{N}{\mathcal A}_{x}((W))+2K^x=\O{\bf r} +{\bf s}\;.
\ee

The tau function on the space ${\Mb}_g^{N,0}$ is defined by
\begin{eqnarray}
\tau(\CC,W)=\la{taudefW}
\Ccal^{2/3}(x) e^{-\f{\pi}{6} \langle\O {{\bf r}},{{\bf r}}\rangle-\f{2\pi \i}{3}\langle{{\bf r}},K^x\rangle}
\left(\f{{\mathcal{W}}(x)}{\prod_{i=1}^{m}E(x,x_i)}\right)^{\frac{g-1}{3N}} 
\prod_{i<j} E(x_i,x_j)^{\f{1}{6N^2}},
\end{eqnarray}
see  \cite{KSZ} for details. 

The tau function (\ref{taudefW}) satisfies the following system of equations with respect to the periods of $w$
(\ref{Ptco}):
\be
\f{\p \log \tau(\CC,W)}{\p \Pt_i}=-\f{1}{2\pi \i N}\int_{\st_i^*}\f{B^{reg}_w}{w}
\la{vartau}\ee
where
\be
{B}^{reg}_w(x)=\left({B}(x,y)-\f{w(x)w(y)}{(\int_y^x w)^2}\right)\Big|_{y=x}\;.
\la{Bregw}
\ee

The tau function  (\ref{taudefW}) has properties similar to those of (\ref{taupfor}):
\begin{itemize}
\item
Under the change (\ref{symtrans}) of a Torelli marking of $\CC$ the tau function (\ref{taudefW}) transforms as follows:
\be
\tau(\CC,W)\to \rho\, \det(C\O+D)\, \tau(\CC,W)\;,
\la{chTor}
\ee
where $\rho$ is a root of unity of order $48(N+1)$. 
\item
$\tau(\CC,\mu W)$ is quasi-homogeneous with respect to the action of $\C^*$:
\be
\tau(\CC,\e W)= \e^\kappa\tau(\CC,W),\quad\e\in\C^*,
\la{homog}
\ee
with
\be
\ka= \f{(2N+1)(g-1)}{6N(N+1)}\;.
\la{homco}
\ee
\end{itemize}

These properties, together with the asymptotics of $\tau(\CC,W)$ near $D_{deg}$ and the components of the Deligne-Mumford boundary,
imply the following expression for the Hodge class on the space $P\overline{\Mb}_g^{N}$ (Theorem 3.9 of \cite{KSZ}):
\be
\l=\f{(g-1)(2N+1)}{6N(N+1)}\psi +\f{1}{12N(N+1)} \delta_{deg}+\f{1}{12}\delta\;.
\la{Hodgeint}
\ee


\section{The divisor class of the universal Hitchin's discriminant}

Consider the following vector bundles on the moduli spaces of curves and their pullbacks to $P\overline{\Mcal}$:
\begin{itemize}
\item
The Hodge vector bundle $\Lambda\to\overline{\M}_g$. The fiber of $\Lambda$ over a smooth curve  $\CC$ is the $g$-dimensional vector space of holomorphic 1-forms (Abelian differentials) on $\CC$.
This bundle naturally lifts to $P\overline{\Mcal}$, and we put $\lambda=c_1(\det\Lambda)$. 

\item
The Hodge vector bundle $\Lh\to\overline{\M}_{\gh}$.The fiber of $\Lh$ over a smooth spectral cover $\Ch$ is the $\gh$-dimensional vector space of holomorphic 1-forms on $\Ch$.
This bundle also lifts to $P\overline{\Mcal}$, and, similarly, we put $\lh=c_1(\det\Lh)$.

\item The tautological line bundle $L$. The line bundle $L$ is associated with the natural action of $\C^*$ on $\Mcal$ by
\be
\dif_k\mapsto \e^k \dif_k \;,\quad \e\in \C^*\,.
\la{rescale}
\ee
Denote by $P\Mcal$ the projectivization of $\Mcal$
with respect to the action (\ref{rescale}). The fibers of the projection $P\M\to\M_g$ are weighted projective spaces 
$\M^{\CC}/\C^*$, where $\M^{\CC}=\bigoplus_{j=1}^n\,H^0(\CC,K_{\CC}^{\otimes j})$, see \eqref{wpr}.
The bundle $L$ extends to the compctification $P\overline{\Mcal}$, and we put $\phi=c_1(L)$.
\end{itemize}
\begin{remark}{\rm
Rigorously speaking, all these objects should be understood in a proper sense (that is, as sheaves on smooth algebraic stacks). However, abusing the language, we will continue calling them vector bundles.} 	
\end{remark} 

If the base curve $\CC$ has nodes, the differentials $\dif_j$ may have poles up to order $j$ at each node. If $\dif_j$ has poles of the maximal order $j$ at the two intersecting branches of $\CC$ with equal or opposite $j$-residues depending on the parity of $j$; see Section 1.1  of \cite{KSZ} or \cite{GruMol} for details. Therefore, the discriminant $W$ can have poles of order up to $n(n-1)$ at the nodes (in case of poles of order $n(n-1)$ the residues must be equal, since $n(n-1)$ is always even).


For a point $(\CC,\{\dif_k\}_{k=1}^n)\in \Mcal$ consider the cyclic $\Z/{N\Z}$-cover $\Ct$ of $\CC$ given by
\eqref{Nco} with $N=n(n-1)$, and consider the decomposition (\ref{decH}) of the homology group $H_1(\Ct,\C)$. Choose a set of $(n(n-1)+1)(2g-2)$
linearly independent cycles 
\be
\st_1,\dots,\st_{2(n^2-n+1)(g-1)}
\la{stdef1}\ee
in the subspace ${\cal S}_1$. As in (\ref{stdefdu}),
consider the  cycles dual to (\ref{stdef}) 
\be
\st_1^*,\dots,\st^*_{2(n^2-n+1)(g-1)},\quad \st_i^*\circ \st_j=\delta_{ij}
\la{stdefdu1}\ee
(they form a basis in the  space ${\cal S}_{n(n-1)-1}$.
Generally speaking, the periods of $w$ with respect to the basis (\ref{stdef1}) {\it do not} provide a coordinate system on 
$\Mcal$ since
$\dim\, \Mcal=(n^2+3)(g-1)+1$ is smaller than $\dim\,{\cal S}_1$ for $N\geq 3$.

The tau function $\tau$ on the space $\Mcal$ can be defined by the system of equations
\be
d{\rm log}\tau(\CC,W)=   -\f{1}{2\pi \i n(n-1)}\sum_{i=1}^{2(n^2-n+1)(g-1)}\left(\int_{\st_i^*}\f{B^{reg}_w}{w} \right)\; d\left(\int_{\st_i}w\right)\;.
\la{deftauW}
\ee 
The solution $\tau(\CC,W)$ of (\ref{deftauW}) is given by the formula (\ref{taudefW}), where $N=n(n-1)$ and
$x_i$, $i=1,\dots,n(n-1)(2g-2)$ are the zeroes of $W$.

Since under the rescaling $\dif_k\mapsto \e^k\dif_k,\;\e\in\C^*$, 
the discriminant $W$ transforms as $W\mapsto \e^{n(n-1)} W$,
the tautological line bundles $\Lcal$ and $L$ associated with the $\C^*$-actions on $\overline{\Mb}_g^{N}$ 
and $\overline{\Mcal}$ respectively are related by $\Lcal\simeq L^{n(n-1)}$, and $\psi=c_1(\Lcal)=n(n-1)c_1(L)=n(n-1)\phi$.

Furthermore, according to formulas (3.13), (3.15) of \cite{KSZ}, the tau function $\tau(\CC,W)$ has the following asymptotics when two zeros  of $W$
(say, $x_1$ and $x_2$) coalesce:
\be
\tau(\CC,W)\sim \left(\int_{x_1}^{x_2}w\right) ^{\frac{1}{(n^2-n+1)(n^2-n+2)}} (1+ o(1))\;.
\la{astau}
\ee

Using the transformation properties (\ref{chTor}) and (\ref{homog}) of $\tau(\CC,W)$ and computing its divisor, we obtain
the following relation in ${\rm Pic}(P\overline{\Mcal})\otimes\Q$
\be
12n(n-1)\l=\f{1}{(n^2-n+1)}\, [PD_W]+\f{2(2n^2-2n+1)(g-1)}{n^2-n+1}\,\psi+n(n-1)\,\delta\;,
\la{HodgeW}
\ee
where $\delta$ is the pullback of the Deligne-Mumford boundary class relative to the projection $P\olMcal\to\olMcal_g$.
Expressing the class of $PD_W$ in terms of $\phi$, $\lambda$ and the boundary class we get
\begin{theorem}\la{PWd}
The class of the (projectivized)  universal Hitchin's discriminant   $PD_W$   defined by (\ref{DWdec}) expresses in terms of the
standard generators of ${\rm Pic}(P\overline{\Mcal})\otimes\Q$ as follows:
\be
\f{1}{n(n-1)}[PD_W]=(n^2-n+1)(12\lambda -\delta)- 2(2n^2-2n+1)(g-1) \phi \;.
\la{formMW}
\ee


\end{theorem}
\section{Divisor $PD_v$ and the Hodge class $\lh$}

There is a natural map $\Mcal\to\Hcal_{\gh}$ to the moduli space of holomorphic 1-forms that sends the point $(\CC,\{\dif_i\}_{i=1}^n)\in \Mcal$ to the point  $(\Ch,v)\in \Hcal_{\gh}$.
Generically, all zeros of the differential $v$ are simple, and there are $2\gh-2$ of them that we denote $y_1,\ldots,y_{2\gh-2}$.
The number of periods of $v$ over the cycles (\ref{sj}) in
the relative homology group
$H_1(\Ch,\{y_1,\ldots,y_{2\gh-2}\})$ is equal to
$4\gh-3=4n^2(g-1)+1$ which is in general greater than $\dim\,\Mcal= (n^2+3)(g-1)+1$.

Consider the set of generators $s_j$ of the relative 
homology group $H_1(\Ch,\{y_1,\ldots,y_{2\gh-2}\})$:
\be
\{s_1,\ldots,s_{4\gh-3}\}=(\ah_1,\dots,\ah_{\gh},\bh_1,\dots,\bh_{\gh},l_2,\dots l_{2\gh-2})\;,
\la{sjv}
\ee
where $l_j$ is a simple path connecting $y_1$ with $y_j$.

The dual system of generators in the homology group  $H_1(\Ch\setminus\{y_1,\ldots,y_{2\gh-2}\})$ is 
\be
\{s_1^*,\ldots,s_{4\gh-3}^*=\{-\bh_1,\dots,-\bh_{\gh},\ah_1,\dots,\ah_{\gh},c_2,\dots,c_{2\gh-2})\;,
\la{sjvstar}
\ee
where $c_j$ is a small positively oriented circle around $y_j$ such that $s_i^*\circ s_j=\delta_{ij}$)


The class of the divisor of zeros of the differential $v$ can be expressed in terms of the Hodge class $\hat{\lambda}=c_1(\hat{\Lambda})$ 
and the classes $\psi$ and $\delta$
using the tau function  (\ref{taupfor}) on the moduli spaces of holomorphic  Abelian differentials 
with simple zeros on the complex curves of genus $\gh$. The tau function $\tau(\Ch,v)$ on the space of spectral covers (\ref{cancov}) is defined by the explicit formula (\ref{taupfor}).





Formula (\ref{sym}) implies that $\tau(\Ch,v)$ transforms like follows under the change of Torelli marking of $\Ch$ given by
$\left(\begin{array}{cc} \hat{C} & \hat{D}\\ \hat{B} & \hat{A}\end{array}\right)\in Sp(2\gh,\Z)$: 
 \be
\tau(\Ch,v)\to \rho\, \tau(\Ch,v)\, \det(\hat{C}\,\Oh+\hat{D})
\la{sym2}
\ee
where $\rho^{24}=1$.
By (\ref{hom}), under the rescaling $v\mapsto\e v,\;\e\in\C^*,$ $\tau(\Ch,v)$ behaves like 
\be
\tau(\Ch,\e v)=\e^{(\gh-1)/4}\tau(\Ch, v)\;.
\la{hom2}
\ee

Notice that when the curve $\CC$ approaches the boundary of $\overline{\M}_g$, 
the cover $\Ch$ approaches a codimension $n-1$ locus $\hat{D}_0$ in the component ${\delta}_0$ of the Deligne-Mumford boundary of $\overline{\M}_{\gh}$.
Then the formulas (\ref{sym2}) and (\ref{hom2}) combined with the asymptotics 
of $\tau(\Ch,v)$ near $\delta_0$ in $\overline{\M}_{\gh}$ (cf. Lemma 7 of \cite{MRL}),
imply the following 
\begin{theorem} 
\la{Pvd}
The class of the (projectivized) divisor $PD_v$ of non-generic (i.e., for $v$ with multiple zeroes) $GL(n)$ spectral covers 
in ${\rm Pic}(P{\overline{\Mcal}})\otimes\Q$ is given by
\be
[PD_v]=24\lh-6(\gh-1)\phi -2n\delta\;.
\la{Dvcl}
\ee
Here 
$\lh$ is the Hodge class of $\overline{\M}_{\gh}$ pulled back to $P{\overline{\Mcal}}$,
$\phi$ is the tautological class associated with the projection $\ol{\Mcal}\to P\ol{\Mcal}$, and $\delta$ is the
pullback to $P\ol{\Mcal}$ of the Deligne-Mumford boundary of $\overline{\M}_{g}$.
\end{theorem}
The proof of the theorem follows almost verbatim the proof of Theorem 2 in \cite{KSZ}

\section{Prym class in ${\rm Pic}(P{\overline{\Mcal}})\otimes \Q$}

Let $y\in \CC$ be a generic point of the projection $\pi:\Ch\to \CC$, i.e. $|\pi^{-1}(y)|=n$. Denote by $\xi$ a local coordinate on $\CC$ in a small neighborhood $U$ of $y$.
Then $\xi$ can be used as a local coordinate on each of the $n$ connected components of the preimage of $U$.

A holomorphic Abelian differential $u$ on $\Ch$ is called a {\em Prym differential} if
\be
\sum_{x\in \pi^{-1}(y)} \frac{u}{d\xi}(x) =0
\ee
for any $y\in \CC$ that is not a branch point of $\Ch$).
Then there is the following decomposition of the space of holomorphic differentials on $\Ch$:
\be
\Omega^1_{\Ch}=\Omega^1_{\CC}\oplus H_{Prym}^{1}(\Ch)\;.
\la{decom}
\ee

If the cover (\ref{cancov}) arises from an $SL(n)$ Hitchin's system (i.e. if $\dif_1=0$), then the sum of solutions of the
equation (\ref{cancov}) is zero. In this case $v\in H_{Prym}^{1}(\Ch)$ is a Prym differential.

The vector bundle on $\Mcal$ with fiber $H_{Prym}^{1}(\Ch)$ over $\Ch$ is called the {\it Prym vector bundle}. The Prym bundle naturally extends to $\overline{\Mcal}$ and descends to the projectivization $P\overline{\Mcal}$. 
The first Chern class of the determinant of the Prym vector bundle is called the {\it Prym class} and is
denoted by $\lambda_P$.

We define the {\it Prym tau function} $\tau_P$ as
\be
\tau_P=\frac{\tau(\Ch,v)}{\tau(\CC,W)}\;.
\la{tauP}
\ee
It may be viewed as a section of a holomorphic line bundle on $P\overline{\Mcal}$. 
Computing its divisor and using Theorems  \ref{PWd} and \ref{Pvd} we get
\begin{corollary}
The Prym class $\lambda_P$
decomposes in ${\rm Pic}(P{\overline{\Mcal}})\otimes \Q$ into a linear combination of the classes $[PD_W]$, $[PD_v]$, the tautological class $\phi$ and the boundary class $\delta$ as follows:
\be
\lambda_P=\f{1}{24}[PD_v]-\f{1}{12n(n-1)(n^2-n+1)}[PD_W]+\f{g-1}{3}\left(n^2-\f{2n^2-2n+1}{2(n^2+n+1)}\right)\phi
+\f{n-1}{12}\delta
\la{Prymf}
\ee
\end{corollary}

\section{$GL(2)$ spectral covers}

The equation (\ref{cancov}) of the spectral cover $\Ch$ in the $GL(2)$ case looks like follows:
\be
v^2+ \dif_1 v+\dif_2=0
\la{cancov1}
\ee
The discriminant is then
$W=\dif_1^2-4\dif_2$,
and the differential $v$ on $\Ch$ is
\be
v=\frac{1}{2}(-\dif_1\pm \sqrt{W})\;,
\la{vGL}
\ee
where the choice of $\sqrt{W}$ is compatible with the involution $\Ch\to\Ch$. Generically, the differential $v$ has $4g-4$ simple zeros
at the branch points (since both $\dif_1$ and $\sqrt{W}$, being lifted to $\Ch$, have simple zeros at the branch points), and $4g-4$ more simple poles elsewhere.

If all $4g-4$ zeros of $W$ are simple, then the genus of $\Ch$ equals to $\gh=4g-3$.
The formula (\ref{HodgeW}) for the class of the divisor $PD_W$ (which in this case coincides with  $PD_W^{(b)}$) takes the form
\be
[PD_W]=72\lambda -20(g-1)\phi -6\,\delta\;,
\la{MW2}
\ee
and for the divisor $PD_v$ by (\ref{Dvcl}) we have 
\be
[PD_v]=24\lh-24(g-1)\phi-4\,\delta\;.
\la{Mv2}
\ee

\section{Open questions}
The following questions arise naturally in connection with the subject of this work.

1.
Using (\ref{Dvcl}) one can  express the class
$[PD_v]$ in terms of the class $\phi$ and the class $(\widehat{\Theta})$ of the 
divisor of zeros of the product  of even theta constants (the ``theta-null") on $\Ch$.
This relation follows from (\ref{Dvcl}) and the expression of $(\widehat{\Theta})$ in terms
of $\hat{\lambda}$ given in Proposition 3.1 of \cite{Teixidor}.
Similarly, by Formula (\ref{formMW}) and Proposition 3.1 of \cite{Teixidor} one can express the class of  divisor  $PD_W$ in terms of the class of the theta-null
on $\CC$, the tautological class $\phi$ and  the boundary class $\delta$.
What kind of relation one can get between the divisor classes of ``theta-nulls" of the cover and of the base?

2. 
The  holomorphic $n(n-1)$-differentials which appear as discriminants of Hitchin's $GL(n)$ covers
are rather special: for a fixed $\CC$ the space of discriminants has dimension $g+(g-1)\sum_{k=2}^n (2k-1)=n^2(g-1)+1$,
while the space of all holomorphic $n(n-1)$-differentials has dimension $(2n(n-1)-1)(g-1)$. How
to distinguish differentials that are discriminants among all holomorphic $n(n-1)$-differentials?

3. What is the connection, if any, between the divisors appearing in this work and the ``critical loci" discussed in the  recent paper
of N.Hitchin \cite{Hitchin3}?

\begin{remark}\rm
This paper was originally published in the volume \cite{JMP} dedicated to the memory of L.~D~.Faddeev.
More recently the paper \cite{Basok} by M.~Basok appeared containing more detailed information about various divisor classes discussed in this paper.
More precisely, for $n\geq 3$ and $g\geq 1$ the following formulas were obtained in 
\cite{Basok} for the classes of the ``caustic", the ``Maxwell stratum" and the boundary components of the universal Hitchin's discriminant  (\ref{DWdec}):
  $$
    [P\overline{D}_W^{(c)}] = n(n-1)(n-2)\Bigl( 12\lambda-\delta - 4(g-1)\phi \Bigr)\;,
    $$
    $$
   [P\overline{D}_W^{(m)}] = \frac{n(n-1)(n-2)(n-3)}{2}\Bigl( 12\lambda-\delta + 4(g-1)\phi \Bigr)\;,
   $$
   $$
  [P\overline{D}_W^{(b)}] = n(n-1)\Bigl( (n+1)(12\lambda-\delta) - 2(g-1)(2n+1)\phi \Bigr)\;.
$$
 
 The expression for the class of the universal Hitchin' discriminant $D_W$ obtained by means of these formulas coincides with our formula (\ref{formMW}).

Another formula derived in \cite{Basok} relates the classes $\lambda$ and $\widehat{\lambda}$:
  \begin{equation*}
    \widehat{\lambda} = n(2n^2-1)\lambda - \frac{n(n-1)(4n+1)(g-1)}{6}\phi - \frac{n(n^2-1)}{6}\delta.
  \end{equation*}
In particular, this formula yields a relation between the classes $[D_W]$ and $[D_v]$ given by the formulas \eqref{formMW} and \eqref{Dvcl}.
\end{remark}

{\bf Acknowledgements.}  The authors thank Mikhail Basok for discussions. D.~K. thanks Marco Bertola, Jacques Hurtubise and Chaya Norton for useful comments.
We thank Michael Baker for spotting several misprints in the first version of this paper.
The work of D.K. was supported in part by the Natural Sciences and Engineering Research Council of Canada grant
RGPIN/3827-2015 and by
the  FQRNT grant "Matrices Al\'eatoires, Processus Stochastiques et Syst\`emes Int\'egrables" (2013--PR--166790). 
The research of Section 3 was supported by the Russian Science Foundation grant 14-21-00035.

\end{document}